\pdfoutput=1
\documentclass[letterpaper,twocolumn,10pt]{article}
\usepackage{usenix,epsfig,endnotes}
\usepackage[linesnumbered,algoruled,boxed,lined]{algorithm2e}
\usepackage{caption,subcaption}
\usepackage{tablefootnote}
\usepackage{fancyvrb}
\usepackage{multirow,array,url}
\usepackage[misc,geometry]{ifsym}
\usepackage{bold-extra}
\usepackage{pifont}
\usepackage[usenames,dvipsnames,svgnames,table]{xcolor}
\usepackage{arydshln}
\usepackage{flushend}
\usepackage{amsmath}
\usepackage{hyperref}

\SetAlFnt{\small}
\SetAlCapFnt{\normalsize}
\SetAlCapNameFnt{\normalsize}

\makeatletter
\g@addto@macro{\UrlBreaks}{\UrlOrds}
\makeatother

\graphicspath{ {./figures/} }

\newcommand{\xmark}{\ding{55}}
\newcommand{\checkmark}{\ding{51}}
\newcommand{\ld}{\texttt{Automatic Lockdown}}
\newcommand{\lds}{\texttt{AutoLock}}
\newcolumntype{C}[1]{>{\centering\let\newline\\\arraybackslash\hspace{0pt}}m{#1}}

\begin{document}

%
%
\title{\Large \bf {\LARGE \texttt{AutoLock}}: Why Cache Attacks on ARM Are Harder Than You Think}
\hypersetup{pdfauthor={Green et al.},pdftitle={AutoLock: Why Cache Attacks on ARM Are Harder Than You Think}}

%
%
\author{
 {\rm Marc Green}\\
 Worcester Polytechnic Institute
 \and
 {\rm \hspace*{-6mm}Leandro Rodrigues-Lima}\hspace*{4mm}\\
 \hspace*{-6mm}Fraunhofer AISEC\hspace*{4mm}
 \and
 {\rm Andreas Zankl\hspace*{9mm}}\\
 Fraunhofer AISEC\hspace*{9mm}
 \and
 {\rm Gorka Irazoqui}\\
 Worcester Polytechnic Institute
 \and
 {\rm Johann Heyszl}\\
 Fraunhofer AISEC
 \and
 {\rm Thomas Eisenbarth}\\
 Worcester Polytechnic Institute
 }
\maketitle

%
%
\begin{abstract}
Attacks on the microarchitecture of modern processors have become a practical threat to security and privacy in desktop and cloud computing. Recently, cache attacks have successfully been demonstrated on ARM based mobile devices, suggesting they are as vulnerable as their desktop or server counterparts. In this work, we show that previous literature might have left an overly pessimistic conclusion of ARM's security as we unveil \lds: an internal performance enhancement found in inclusive cache levels of ARM processors that adversely affects \texttt{Evict+Time}, \texttt{Prime+Probe}, and \texttt{Evict+Reload} attacks. \lds's presence on system-on-chips (SoCs) is not publicly documented, yet knowing that it is implemented is vital to correctly assess the risk of cache attacks. We therefore provide a detailed description of the feature and propose three ways to detect its presence on actual SoCs. We illustrate how \lds~impedes cross-core cache evictions, but show that its effect can also be compensated in a practical attack. Our findings highlight the intricacies of cache attacks on ARM and suggest that a fair and comprehensive vulnerability assessment requires an in-depth understanding of ARM's cache architectures and rigorous testing across a broad range of ARM based devices.
\end{abstract}

%
%
\section{Introduction}
\label{sec:intro}
The rapid growth of mobile computing illustrates the continually increasing role of digital services in our daily lives. As more and more information is processed digitally, data privacy and security are of utmost importance. One of the threats known today aims directly at the fabric of digital computing. Attacks on processors and their microarchitecture exploit the very core that handles our data. In particular, processor caches have been exploited to retrieve sensitive information across logic boundaries established by operating systems and hypervisors. As caches speed up the access to data and instructions, timing measurements allow an adversary to infer the activity of other applications and the data processed by them. In fact, cache attacks have been demonstrated in multiple scenarios in which our personal data is processed, e.g., web browsing~\cite{OrenEtAl2015} or cloud computing~\cite{InciEtAl2016,ZhangEtAl2014}. These attacks have severe security implications, as they can recover sensitive information such as passwords, cryptographic keys, and private user behavior. The majority of attacks have been demonstrated on classic desktop and server hardware~\cite{GullaschEtAl2011,IrazoquiEtAlAsia2015,OsvikEtAl2006,YaromFalkner2014}, and with Intel's market share for server processors being over 98\%~\cite{Kay2014}, their platforms have been targeted most frequently.

With mobile usage skyrocketing, the feasibility of cache attacks on smartphone and IoT processors -- which are predominantly ARM-based -- has become a relevant issue. Attacks that rely on the existence of a cache flush instruction, i.e., \texttt{Flush+Reload}~\cite{YaromFalkner2014} and \texttt{Flush+Flush}~\cite{GrussEtAl2016a}, work efficiently across a broad range of x86 processors, but have limited applicability on ARM devices. In general, cache flush instructions serve the legitimate purpose of manually maintaining coherency, e.g., for memory mapped input-output or self-modifying code. On any x86 processor implementing the SSE2 instruction set extension, this flush instruction is available from all privilege levels as \texttt{clflush}. A similar instruction was introduced for ARM processors only in the most recent architecture version, ARMv8. In contrast to \texttt{clflush}, it must be specifically enabled to be accessible from userspace. This leaves a significant number of ARM processors without a cache flush instruction.

For all processors with a disabled flush instruction or an earlier architecture version, e.g., ARMv7, only eviction based cache attacks can be deployed. In particular, these attacks are \texttt{Evict+Time}~\cite{OsvikEtAl2006}, \texttt{Prime+Probe}~\cite{OsvikEtAl2006}, and \texttt{Evict+Reload}~\cite{GrussEtAl2015}. On multi-core systems, they target the last-level cache (LLC) to succeed regardless of which core a victim process is running on. This requires the LLC to be inclusive, i.e., to always contain the contents of all core-private cache levels. On Intel processors, the entire cache hierarchy fulfills the inclusiveness property and is therefore a viable target for eviction based attacks. ARM devices, on the contrary, implement inclusive and non-inclusive caches alike. Both properties can co-exist, even in the same cache hierarchy. This renders eviction based attacks to be practicable only on a limited number of devices, in particular those that implement inclusive last-level caches. Yet, our findings show that an internal performance enhancement in inclusive last-level caches, dubbed \lds, can still impede eviction based cache attacks. In short, \lds~prevents a processor core from evicting a cache line from an inclusive last-level cache, if said line is allocated in any of the other cores' private cache levels. This inhibits cross-core LLC evictions, a key requirement for practical \texttt{Evict+Time}, \texttt{Prime+Probe}, and \texttt{Evict+Reload} attacks on multi-core systems, and further limits the number of ARM based attack targets in practice.

In literature, Lipp et al.~\cite{LippEtAl2016}, Zhang et al.~\cite{ZhangEtAl2016a}, and Zhang et al.~\cite{ZhangEtAl2016} confirmed the general feasibility of flush and eviction based cache attacks from unprivileged code on ARM processors. Given the lack of flush instructions on a large selection of ARM devices and the deployment of non-inclusive LLCs or inclusive LLCs implementing \lds, the authors might have left an overly pessimistic conclusion of ARM's security against cache attacks. In addition, ARM's highly flexible licensing ecosystem creates a heterogeneous market of system-on-chips (SoCs) that can exhibit significant differences in their microarchitectural implementations. Demme et al.~\cite{Demme:2012:SVF:2337159.2337172} illustrate that already small changes to the cache architecture can have considerable impact on the side-channel vulnerability of the processor. Consequently, judging the true impact of cache attacks on a broad range of ARM based platforms remains to be a challenge. This work adds another step in this process. It is a contribution to an in-depth understanding of microarchitectural features on ARM in general and an extension to our current knowledge of cache implementations in particular.

\subsection{Contribution}
This work unveils \lds, an internal and undocumented performance enhancement feature found in inclusive cache levels on ARM processors. It prevents cross-core evictions of cache lines from inclusive last-level caches, if said lines are allocated in any of the core-private cache levels. Consequently, it has a direct and fundamentally adverse effect on all eviction based cache attacks launched from unprivileged code on multi-core systems. Understanding \lds~and determining its existence on a given system-on-chip is vital to assess the SoC's vulnerability to those attacks. Yet, neither technical reference manuals (TRMs) nor any other public product documentation by ARM mention \lds. We therefore provide a detailed description of the feature and propose three methodologies to test for it: using a hardware debugging probe, reading the performance monitoring unit (PMU), and conducting simple cache-timing measurements. Each test strategy has different requirements and reliability; having multiple of them is vital to test for \lds~under any circumstances. With the proposed test suite, we verify \lds~on ARM Cortex-A7, A15, A53, and A57 processors. As \lds~is likely implemented on a larger number of ARM processors, we discuss its general implications and how our results relate to previous literature.

Despite its adverse effect on eviction based cache attacks, the impact of \lds~can be reduced. We discuss generic circumvention strategies and execute the attack by Irazoqui et al.~\cite{waitaminute} in a practical cross-core \texttt{Evict+Reload} scenario on a Cortex-A15 implementing \lds. We successfully recover the secret key from a table based implementation of AES and show that attacks can tolerate \lds~if multiple cache lines are exploitable. Furthermore, the presented circumvention strategies implicitly facilitate cross-core eviction based attacks also on non-inclusive caches. This is because in the context of cross-core LLC evictions, inclusive last-level caches with \lds~behave identically to non-inclusive ones. In summary, our main contributions are:

\begin{itemize}
  \item the disclosure and description of \lds, an undocumented and previously unknown cache implementation feature with adverse impact on practical eviction based cache attacks on ARM devices,
  \item a comprehensive test suite to determine the existence of \lds~on actual devices, as its presence is not documented publicly,
  \item a discussion of \lds's implications and its relation to previous literature demonstrating cache attacks on ARM, and
  \item a set of strategies to circumvent \lds~together with a practical demonstration of a cross-core \texttt{Evict+Reload} attack on a multi-core SoC implementing \lds.
\end{itemize}
  
The rest of this paper is organized as follows. Section~\ref{sec:imlockdown} describes \lds. A theoretical methodology to test for it is presented in Section~\ref{sec:howtotest}. We evaluate SoCs for \lds~in Section~\ref{sec:soceval} and address how recent literature relates to it in Section~\ref{sec:relwork}. The implications of \lds~are discussed in Section~\ref{sec:impli}. Circumvention strategies together with a practical cross-core attack are presented in Section~\ref{sec:circum}. We conclude in Section~\ref{sec:concl}.

\section{{\large \textbf{\lds}:} Transparent Lockdown of Cache Lines in Inclusive Cache Levels}
\label{sec:imlockdown}
Processor caches can be organized in levels that build up a hierarchy. Higher levels have small capacities and fast response times. \textit{L1} typically refers to the highest level. In contrast, lower levels have increased capacities and response times. The lowest cache level is often referred to as the last-level cache, or \textit{LLC}. Data and instructions brought into cache reside on cache lines, the smallest logical storage unit. In set-associative caches, lines are grouped into sets of fixed size. The number of lines or \textit{ways} per cache set is called the \textit{associativity} of the cache level. It can be different for every level. Whether a cache level can hold copies of cache lines stored in other levels is mainly defined by the inclusiveness property. If a cache level $x$ is \textit{inclusive} with respect to a higher level $y$, then all valid cache lines contained in $y$ must also be contained in $x$. If $x$ is \textit{exclusive} with respect to $y$, valid lines in $y$ must not be contained in $x$. If any combination is possible, the cache is said to be \textit{non-inclusive}.

Inclusive caches enforce two rules. If a cache line is brought into a higher cache level, a copy of the line must be stored in the inclusive lower level. On multi-core processors, this offers a substantial performance advantage to coherency protocols, because determining whether a line is stored anywhere in the hierarchy is achieved by simply looking into the LLC. Vice versa, if a line is evicted from the lower level, any copy in the higher levels must subsequently be evicted as well. This is an implicit consequence of the inclusiveness property that has been successfully exploited in cross-core cache attacks that target inclusive LLCs~\cite{InciEtAl2016,IrazoquiEtAl2015,LiuEtAl2015}.

Evictions in higher cache levels to maintain inclusiveness can add substantial performance penalties in practice. In a patent publication by Williamson and ARM Ltd., the authors propose a mechanism that protects a given line in an inclusive cache level from eviction, if any higher cache level holds a copy of the line~\cite{Williamson2012}. An \textit{indicator storage element} that is integrated into the inclusive cache level tracks which lines are stored in higher levels. The element can be realized with a set of \textit{indicator} or \textit{inclusion} bits per cache line, or a tag directory. If an indicator is set, the corresponding line is protected. This mechanism therefore prevents said performance penalties, because subsequent evictions in higher cache levels are prohibited. We refer to this transparent protection of cache lines in the LLC as \ld~or \lds.

\begin{figure}[t!]
	\centering                            
    \includegraphics[width=0.47\textwidth, trim=.5cm 1cm .5cm 2cm, clip]{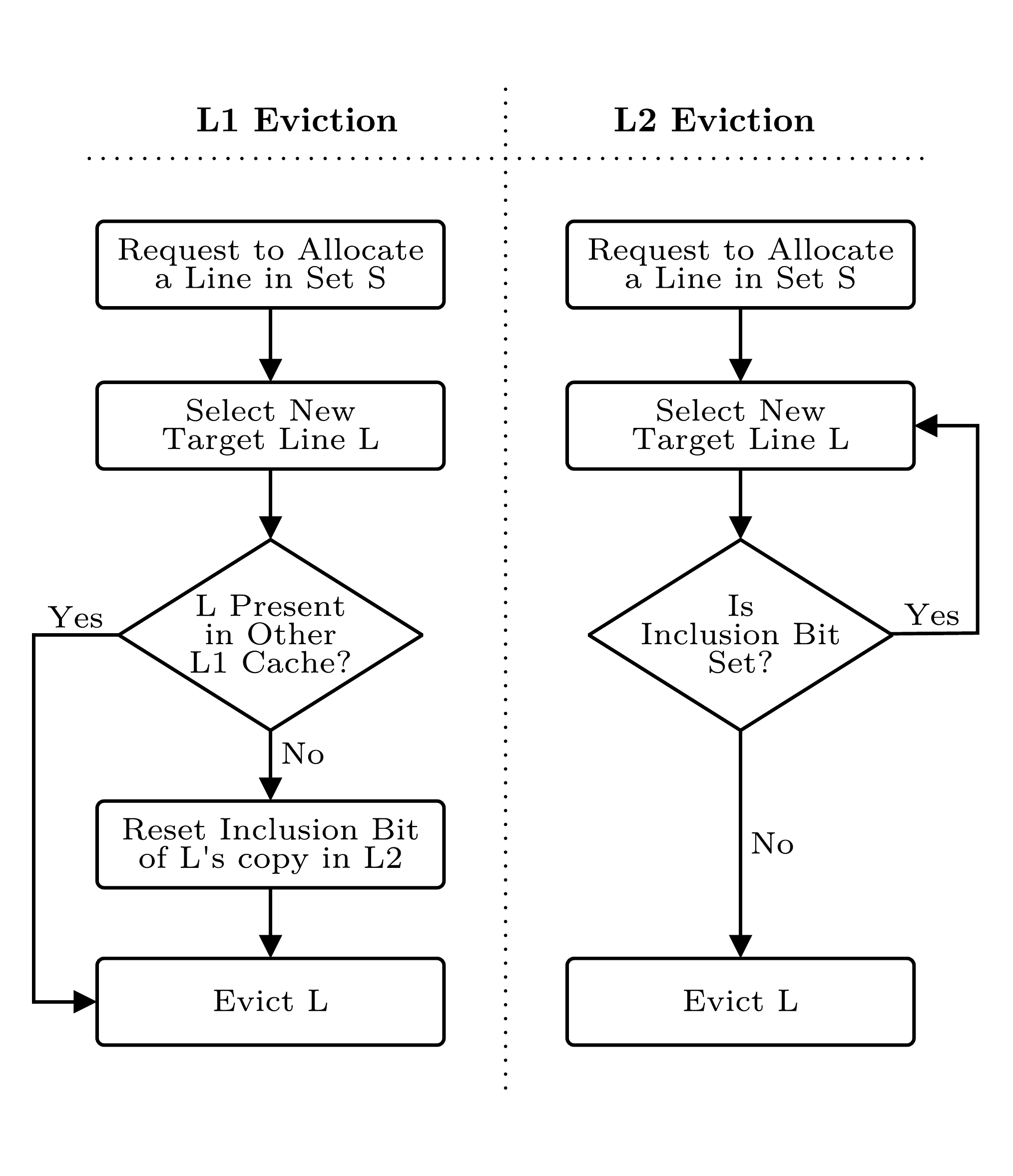}
    \caption{Simplified example of evicting a cache line from level 1 (left) and level 2 (right) cache sets. The level 2 cache is inclusive with respect to level 1 and implements \lds.}
    \label{fig:patentflow}
\end{figure}

The impact of \lds~during eviction is illustrated in Figure~\ref{fig:patentflow}. For simplicity, the illustration is based on a two-level cache hierarchy: core-private L1 caches and a shared inclusive last-level cache (L2) with one inclusion bit per cache line. $S$ and $L$ are placeholders for any available cache set and line in the two cache levels. The left side of the figure shows how a cache line is evicted in L1. First, an allocation request for set $S$ is received. Then, a target line $L$ is selected by the replacement algorithm for eviction. If a copy of $L$ is present in any other of the core-private L1 caches, it can immediately be evicted without updating the inclusion bit in L2. Because other L1 copies exist, the bit does not need to be changed. If no other L1 cache holds a copy of $L$, the inclusion bit must be reset in L2, which unlocks the copy of $L$ in L2. After the bit is reset, $L$ is evicted from L1.

Similarly, an allocation request in L2 triggers the replacement algorithm to select a line in set $S$ for eviction. Before $L$ is evicted in L2, its inclusion bit is checked. If a copy of $L$ exists in any L1 cache, the replacement algorithm is called again to select another target line. This is repeated until one line is found whose inclusion bit is not set. This line is then evicted to allow the allocation of a new one.

If the number of ways in the inclusive lower cache level, $W_{l}$, is higher than the sum of ways in all higher cache levels, i.e., $W_{h,sum} = \sum^{N}_{i=1}{W_{h,i}}$, it can be guaranteed that at least one line is always selectable for eviction. If $W_{l} = W_{h,sum}$, all lines of a set in the lower cache level can be auto-locked. In this case, the patent proposes to fall back to the previous behavior, i.e., evict all copies of a line from higher level caches. This unlocks the line in the lower cache level and subsequently enables its eviction. If the number of ways in the lower cache level is further reduced, such that $W_{l} < W_{h,sum}$, additional measures must be taken to implement inclusiveness and \ld. While not impossible per se, this case is not covered by the patent authors.

If an inclusive LLC with \lds~is targeted in a cache attack, the adversary is not able to evict a target's data or instructions from the LLC, as long as they are contained in the target's core-private cache. In theory, the adversary can only succeed, if the scenario is reduced to a same-core attack. Then, it is possible once again to directly evict data and instructions from core-private caches. Note that the same attack limitation is encountered on systems with non-inclusive last-level caches, because cache lines are allowed to reside in higher levels without being stored in lower levels. In both cases, \lds~and non-inclusive LLC, the attacks do not work cross-core because the attacking core cannot influence the target core's private cache. Note that it is possible, and indeed common on ARM processors, that there are separate L1 caches for instructions and data and that the LLC is inclusive with respect to one of them, but non-inclusive with respect to the other. In Section~\ref{sec:circum}, we discuss possible strategies to circumvent \lds~and re-enable cross-core cache attacks. Because of said similarities, those strategies also enable cross-core attacks on non-inclusive LLCs.

Distinct from \ld, there exists \emph{programmable lockdown} in some ARM processors. Regardless of inclusiveness, it allows the user to explicitly lock and unlock cache lines by writing to registers of the cache controller. This has the same effect as \lds, i.e., the locked cache line will not be evicted until it is unlocked. In contrast, however, programmable lockdown must be actively requested by a typically privileged user. Of the four Cortex-A processors we study in this paper, the technical reference manuals do not mention programmable lockdown for any of them~\cite{A15TRM,A53TRM,A57TRM,A7TRM}. \ld, however, is found in all of them.

\section{How to Test for {\large \textbf{\lds}}}
\label{sec:howtotest}
\lds~is neither mentioned in ARM's architecture reference manuals~\cite{ARM2014,ARMLimited2014} nor in the technical reference manuals of the Cortex-A cores considered in this work~\cite{A15TRM,A53TRM,A57TRM,A7TRM}. To the best of our knowledge, it is not publicly documented other than in patent form~\cite{Williamson2012}. Based on official information, it is therefore impossible to determine which Cortex-A or thereto compliant processor cores implement \lds, let alone whether an actual system-on-chip features it. The presence of \lds, however, is crucial to assess the risk of cache attacks, in particular those that rely on cross-core evictions in the LLC. We therefore propose the following test methodology to determine the existence of \lds. On any device under test, two processes are spawned on distinct cores of the processor implementing an inclusive last-level cache. The first process allocates a cache line in the private cache level of the core it is running on. This allocation is done with a simple memory access. The inclusiveness property ensures that a copy of the cache line must also be allocated in the last-level cache. The second process then tries to evict the line from LLC by filling the cache set the line is contained in with dummy data. If the cache line remains in the LLC after the cross-core eviction, the test concludes that \lds~is implemented. This test strategy requires that the eviction itself works reliably. Otherwise, false positives are encountered. The following section describes the eviction and how to ensure its proper function.

\subsection{Cache Eviction}
In order to evict a cache line from LLC, we implement the method described by Gruss et al.~\cite{GrussEtAl2016b} and Lipp et al.~\cite{LippEtAl2016}. Assume that an address $T$ is stored on line $L$ in cache set $S$. In order to evict $L$ from $S$, one has to access a number of addresses distinct from $T$ that all map to $S$. These memory accesses fill up $S$ and eventually remove $L$ from the set. The addresses that are accessed in this process are said to be \textit{set-congruent} to $T$. They are collected in the eviction set $C$. Whenever $C$ is accessed, $T$ is forced out of the set $S$. The sequence of accesses to addresses in $C$ is referred to as the \textit{eviction strategy}.

\begin{algorithm}[b]
\KwIn{\\
 $C$\,...\,list of set-congruent addresses}
 \vspace{2mm}
 \For(\tcc*[f]{\# of windows}){i = 0..N-1}{
   \For(\tcc*[f]{\# reps/window}){j = 0..A-1}{
     \For(\tcc*[f]{\# addrs/window}){k = 0..D-1}{access$\left(C[i+k]\right)$;}}}
  \caption{Sliding window eviction of the form \texttt{N-A-D}~\cite{GrussEtAl2016b,LippEtAl2016}.
 \label{alg:evict_strat}}
\end{algorithm}

The eviction method proposed by Gruss et al.~\cite{GrussEtAl2016b} is shown in Algorithm~\ref{alg:evict_strat}. The idea is to always access a subset of the addresses in $C$ for a number of repetitions, then replace one address in the subset with a new one, and repeat. This essentially yields a window that slides over all available set-congruent addresses. We therefore refer to this method as \textit{sliding window eviction}. In the algorithm, $N$ denotes the total number of generated windows, $A$ defines the repetitions per window, and $D$ denotes the number of addresses per window. The required size of $C$ is given by the sum of $N$ and $D$. The final eviction strategy is then written as the triple \texttt{N-A-D}. The strategy \texttt{23-4-2}, for example, comprises 23 total windows, each iterated 4 consecutive times and containing 2 addresses. Lipp et al.~\cite{LippEtAl2016} demonstrate that sliding window eviction can successfully be applied to ARM processors.

The parameters \texttt{N-A-D} must be determined once for each processor. This is done by creating a list of set-congruent addresses $C$ and exhaustively iterating over multiple choices of $N$, $A$, and $D$. By continuously checking the success of the eviction, the strategy with the least number of memory accesses that still provides reliable eviction can be determined. Generating the list of set-congruent addresses $C$ requires access to physical address information. This is because the last-level caches on our test devices use bits of the physical address as the index to the cache sets. If the parameter search for $N$, $A$, and $D$ is done in a bare-metal setting, physical address information is directly available. Operating systems typically employ virtual addresses that must be translated to physical ones. Applications on Linux, for instance, can consult the file \texttt{/proc/[pid]/pagemap} to translate virtual addresses~\cite{LippEtAl2016}. Although accessing the pagemap is efficient, access to it can be limited to privileged code or deactivated permanently. Alternatively, huge pages reveal sufficient bits of the physical address to derive the corresponding cache set~\cite{IrazoquiEtAl2015}. To find addresses set-congruent to $T$, new memory is allocated and the containing addresses are compared to $T$. If the least significant address bits match while the most significant bits differ, the address will map to $T$'s cache set but will be placed on a different line within the set. If access to physical address information is entirely prohibited, timing measurements can still be used to find set-congruent addresses~\cite{OrenEtAl2015}.

\begin{table}[t!]
	\centering
	\caption{Number of inclusive L1 and L2 ways, and eviction strategy (\texttt{N-A-D}) for the ARM and ARM-compliant processors of the test devices.
	}
	\resizebox{!}{1.1cm}{%
	\label{tab:set_assoc}
\begin{tabular}{C{2.25cm}|C{1.75cm}|C{1.75cm}|C{1.75cm}}\hline
	\textbf{Processor} & \textbf{L1 Ways} & \textbf{L2 Ways} & \textbf{N-A-D}\\
	\hline
	Cortex-A7  & 2 (Instr.) &  8 & 23-4-2 \\
	Cortex-A15 & 2 (Data)   & 16 & 36-6-2 \\
  	Cortex-A53 & 2 (Instr.) & 16 & 25-2-6 \\
	Cortex-A57 & 2 (Data)   & 16 & 30-4-6 \\
	Krait 450\endnote{To the best of our knowledge, there is no official public documentation for the Qualcomm Krait 450. Table~\ref{tab:set_assoc} lists the results of our experiments as well as statements from online articles~\cite{kraitarch}.} & 4 (Data) & 8 & 50-1-1 \\
	\hline
\end{tabular}}
\end{table}

Once $C$ is filled with addresses, they are accessed according to Algorithm~\ref{alg:evict_strat}. If a processor core implements separated data and instruction caches, the manner in which a set-congruent address ought to be accessed differs. Data addresses can be accessed by loading their content to a register with the \texttt{LDR} assembly instruction. Instruction addresses can be accessed by executing a \emph{branch} instruction that jumps to it. When determining \texttt{N-A-D} on devices that might implement \lds, all memory accesses to $T$ and all evictions of it must be performed on a single core. This ensures that \lds~is not interfering with the parameter search. Once the eviction with a triple \texttt{N-A-D} works reliably in the same-core setting, the \lds~tests can be commenced.

\subsection{{\large \textbf{\lds}} Tests}
\lds~is an internal feature of cache architectures on ARM. Its presence on a processor has to be determined only once, as currently nothing indicates that it can be en- or disabled from software. In the subsequent sections we propose three tests that have been designed to prove or disprove the existence of \lds.  All of them follow the general methodology of determining the success of a cross-core eviction strategy that is known to succeed in the same-core scenario. For simplicity, all tests are explained in a dual-core setting. For a system with more processor cores, each test can either be repeated multiple times or extended in order to determine the presence of \lds~simultaneously on all but one core. In the dual-core setting, core $0$ is accessing the target address $T$ and core $1$ is trying to evict it by using the eviction set $C$ and the processor specific eviction parameters \texttt{N-A-D}. Table~\ref{tab:set_assoc} contains the parameters for the processors considered in this work. For all tests, both $T$ and $C$ are listed as inputs required for eviction, while the triple \texttt{N-A-D} is assumed to be correctly set according to Table~\ref{tab:set_assoc}.

\subsubsection{Hardware Debugger}
The first method to test for \lds~is through the usage of a hardware debugger. It allows to halt a processor at will and directly monitor the cache content. A breakpoint is inserted after the eviction of $T$ and the contents of the caches are analyzed. Through this visual inspection, it is possible to determine with very high confidence whether or not $T$ remains in the cache after the eviction strategy is run. Given an inclusive LLC, it is sufficient to confirm that $T$ either remains in L1 or in L2 to prove that \lds~is present. Algorithm~\ref{alg:incl_lockdown} outlines this test.

\DontPrintSemicolon 
\begin{algorithm}[h]
\KwIn{\\
 $T$\,...\,target address\\
 $C$\,...\,corresponding eviction set}
 \vspace{2mm}
 
 Core $0$ brings $T$ to L1 and L2. 
 
 Core $1$ runs eviction strategy using $C$.
 
 Halt processor and inspect caches.
 
 \lIf{$T$ in L1 of Core $0$ or L2}{\lds~ is present}\lElse{\lds~ is \underline{not} present}
 
 \caption{Hardware Debugger Test}
 \label{alg:incl_lockdown}
\end{algorithm}

The hardware debugger test requires a debugging unit, a target platform that supports it, and physical access to the target device. We use the DSTREAM debugging unit~\cite{ARMLimited2017} in combination with the ARM DS-5 development studio to conduct it. Although DSTREAM only supports a limited number of SoCs\endnote{The list of devices supported by DSTREAM can be retrieved from ARM's website at~\url{https://developer.arm.com/products/software-development-tools/ds-5-development-studio/resources/supported-devices}.}, the test described in this section can also be run with other debugging hardware.

\subsubsection{Performance Monitoring Unit (PMU)}
The second test utilizes the performance monitoring unit, which can count the occurrence of microarchitectural events in a processor. The PMU of ARMv7- and ARMv8-compliant processors can be configured to count the number of accesses (hit or miss) to the last-level cache. The corresponding event is defined under the ID \texttt{0x16} in the architectural reference manuals~\cite{ARM2014,ARMLimited2014}. The difference of the access counts before and after reloading the target address $T$ indicates whether the reload fetched the address from the L1 or the L2 cache. A fetch from L1 indicates that the eviction strategy failed and suggests that \lds~is implemented. If the eviction strategy is successful, the target address has to be fetched from external memory. Before querying the slow external memory, the L2 cache is accessed and checked for the target address. This access is counted and indicates that \lds~is not implemented. To determine this extra access to the L2, a reference value $R$ must be obtained before the test. This is done by reading the L2 access counter for a reload with no previous run of the eviction strategy, which guarantees a fetch from core-private cache. The PMU test can the be conducted as outlined in Algorithm~\ref{alg:pmu_test}.

\begin{algorithm}[h]
\KwIn{\\
 $T$\,...\,target address\\
 $C$\,...\,corresponding eviction set\\
 $R$\,...\,L2 access reference count}
 \vspace{2mm}
 
 Core $0$ brings $T$ to L1 and L2.
 
 Core $1$ runs eviction strategy using $C$.\label{alg:pmu_stepE}
 
 Save PMU count of L2 accesses.
 
 Core $0$ reloads $T$.
 
 Save PMU count again and calculate difference $d$.\label{alg:pmu_stepF}
 
 \lIf{$d \approx R$}{\lds~is present}
 \lElse{\lds~ is \underline{not} present}
 
 \caption{PMU Test}
 \label{alg:pmu_test}
\end{algorithm}

This test requires access to the PMU, which on ARM is typically limited to privileged code, unless otherwise configured. Some operating systems, however, allow userspace applications to access hardware performance events. On Linux, for instance, the \texttt{perf} subsystem of the kernel provides this access via the \texttt{perf\_event\_open} system call. In general, the PMU test can be used when the target processor is not supported by a hardware debugger, or if physical access to the device is not given. Since PMU event counts can be affected by system activity unrelated to the \lds~test, it is recommended to repeat the experiment multiple times. The best results can be obtained in a bare-metal setting, where the test code is executed without a full-scale operating system.

\subsubsection{Cache-timing Measurements}
The third experiment uses timing measurements to infer from where in the memory hierarchy the target address $T$ is reloaded after the supposed eviction. If external memory access times are known, the reload time of $T$ can indicate whether \lds~is implemented or not. If after the eviction strategy the reload time is smaller than what is expected for an external memory access, the target address is likely fetched from cache, thus indicating \lds. If the reload time is equal to an external memory access, the eviction strategy was successful and \lds~is likely not present. This test approach is outlined in Algorithm~\ref{alg:timing_lockdown}.

\begin{algorithm}[h]
\KwIn{\\
 $T$\,...\,target address\\
 $C$\,...\,corresponding eviction set\\
 $M$\,...\,external memory access time}
 \vspace{2mm}
 
 Core $0$ brings $T$ to L1 and L2.
 
 Core $1$ runs eviction strategy using $C$.\label{alg:timing_evict}
 
 Core $0$ reloads $T$ and measures reload time $t$.
 
 \lIf{$t < M$}{\lds~ is present}
 
 \lElse{\lds~ is \underline{not} present}
 
 \caption{Cache-timing Test}
 \label{alg:timing_lockdown}
\end{algorithm}

This test has no further requirements other than running code on the system from userspace and having access to a sufficiently accurate timing source. Commonly used timing sources include hardware based time-stamp counters (\texttt{PMCCNTR} for ARM), the \texttt{perf} subsystem of Linux, the POSIX \texttt{clock\_gettime()} function, and a custom thread based timer. If available, a hardware based time-stamp counter is preferred due to its high precision. Further discussions about timing sources can be found in the work by Lipp et al.~\cite{LippEtAl2016} and Zhang et al.~\cite{ZhangEtAl2016}. Similar to PMU event counts, timing measurements can significantly be affected by noise. It is therefore advisable to repeat the proposed test multiple times to get a robust conclusion about whether address $T$ is fetched from cache or external memory.

\section{Finding {\large \textbf{\lds}} in Existing SoCs}
\label{sec:soceval}
In this work, we evaluate the presence of \lds~on four test devices and their corresponding system-on-chips. They are illustrated in Table~\ref{tab:hardware}. The Samsung Exynos 5422 and the ARM Juno r0 SoCs feature two processors with multiple cores each. They are so-called ARM big.LITTLE platforms, on which a powerful processor is paired with an energy efficient one. Together with the Samsung Exynos 5250, these SoCs are part of dedicated development boards or single-board computers. In contrast, the Qualcomm Snapdragon 805 is part of an off-the-shelf mobile phone. In total, the four test devices comprise five different processors: the ARM Cortex-A7, A15, A53, A57, and the Qualcomm Krait 450. Table~\ref{tab:set_assoc} provides more details about them. It shows the number of ways in L1 and L2 caches, and the eviction strategy parameters for all of them. The illustrated processors implement separate L1 instruction and data caches. The number of L1 ways is given only for the side which the L2 cache is inclusive to. The LLCs on the Cortex-A7 and A53 are inclusive to the L1 instruction caches, while the LLCs on the Cortex-A15, A57, and the Krait 450 are inclusive to the L1 data caches. Although the documentation does not explicitly state that the L2 cache on the Cortex-A53 is instruction-inclusive~\cite{A53TRM}, our practical experiments strongly suggest that it is. In addition, the lead architect of the A53 confirmed it in an interview~\cite{A53interview}.

The tests on the Cortex-A processors are initially done in a bare-metal setting. The lack of an operating system eliminates interfering cache usage from system processes, thereby significantly reducing noise. The experiments are then repeated on Linux for verification. The experiments on the Krait 450 are conducted on Linux only. For each processor, we verify that the eviction parameters listed in Table~\ref{tab:set_assoc} can successfully evict cache lines in a same-core setting. More precisely, we verify successful eviction when evicting data cache lines using data addresses, and when evicting instruction cache lines using instruction addresses. We then test for \lds~in the cross-core case with the experiments proposed in the previous section.

\subsection{Test Results}
The subsequent sections present the results for all test methodologies described in Section~\ref{sec:howtotest}. Along with the conclusions about the presence of \lds~on the test devices, details about the practical execution of the experiments are discussed.

\subsubsection{Hardware Debugger}
The SoCs on the ARM Juno and the Arndale development boards are the only ones among the test devices that are supported by DSTREAM. It is therefore possible to visually inspect the L1 caches of the Cortex-A15, A53, and A57 processors, and the L2 caches of the A15 and A57. A hardware limitation of the Cortex-A53 on the ARM Juno board prevents the visual inspection of its L2 cache. To still test for \lds~on the A53, we leverage the inclusiveness property to surmise L2 contents. According to Algorithm~\ref{alg:incl_lockdown}, \lds~can still be concluded, if the target address is contained in the core-private cache of core $0$. This is derived from the inclusiveness property of the L2 cache. 

To conduct the tests, we connect each supported board, in turn, to the DSTREAM and use breakpoints to temporarily halt program execution after the eviction algorithm is run. When halted, we use the Cache View of the DS-5 development studio to visually determine if the target cache line is present in the respective caches. For the A53, we infer the contents of the L2 based on the inclusive L1. We ran the experiments several times on the A15, A53, and A57 processors. All trials indicate each processor's inclusive cache implements \lds.

\begin{table}[t!]
	\centering
	\caption{Platforms used for the evaluation of~\lds.
	}
	\label{tab:hardware}
	
\scalebox{0.84}{
\begin{tabular}{c|c|l}\hline
	\textbf{Device}             & \textbf{System-on-Chip}              & \multicolumn{1}{c}{\textbf{Processor}} \\ \hline
    Arndale                     & Samsung Exynos 5250                  & 2x Cortex-A15    \\ \hdashline[0.5pt/2pt]
  	\multirow{2}{*}{ODROID-XU4} & \multirow{2}{*}{Samsung Exynos 5422} & 4x Cortex-A7     \\
  	                            &                                      & 4x Cortex-A15    \\ \hdashline[0.5pt/2pt]
	\multirow{2}{*}{ARM Juno}   & \multirow{2}{*}{ARM Juno r0}         & 4x Cortex-A53    \\
	                            &                                      & 2x Cortex-A57    \\ \hdashline[0.5pt/2pt]
	Nexus 6                     & Qualcomm Snapdragon 805              & 4x Krait 450     \\
	\hline
\end{tabular}
}
\end{table}

\subsubsection{Performance Monitoring Unit (PMU)}
To verify to results of the Cortex-A53, we conduct the experiment described in Algorithm~\ref{alg:pmu_test} with it. The PMU is configured to count accesses to the L2 cache. We then execute a target instruction on core $0$ and run a \texttt{25-2-6} eviction strategy on core $1$. Before and after reloading the target instruction, we insert 10 \texttt{NOP} instructions. This reduces the effects of pipelining, as the A53 has an 8-stage pipeline. To ensure we only measure exactly the reload of the target instruction, we execute a \texttt{DSB} and \texttt{ISB} instruction before each set of \texttt{NOP}s. These instructions function as \emph{memory barriers}, guaranteeing that memory access instructions will execute sequentially. This is necessary because the ARM architecture allows memory accesses to be reordered to optimize performance.

As a result, we observe that reloading the target instruction after executing the eviction algorithm causes no additional L2 access. This indicates that the eviction failed and the reload was served from L1. If the eviction had succeeded, the event counter would have been incremented by the L2 cache miss. We ran the experiment multiple times and observed consistent results in each trial. This confirms the presence of \lds~on the Cortex-A53.

\subsubsection{Cache-timing Measurements}
\begin{figure}[t!]
  \includegraphics[width=.47\textwidth]{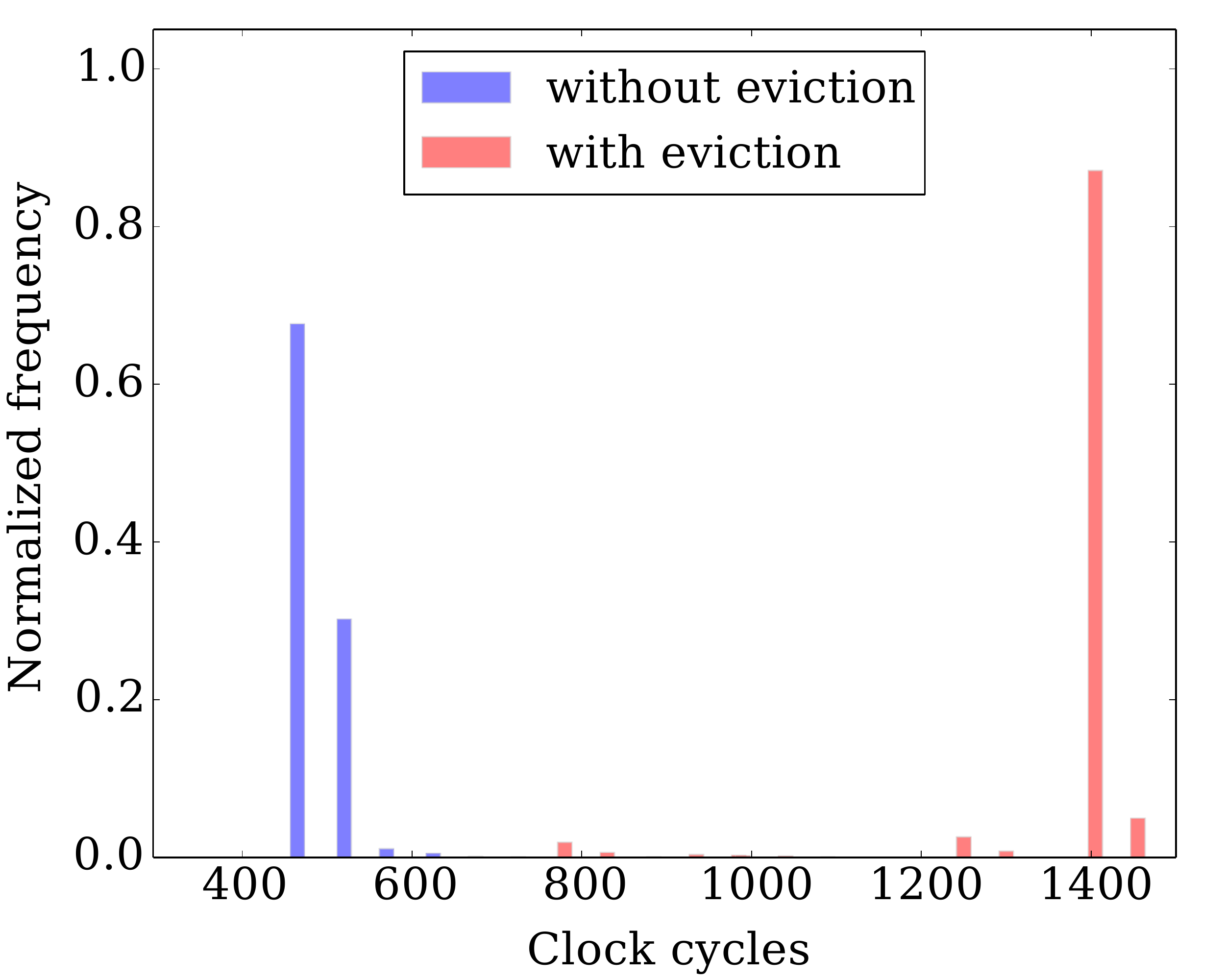}
  \caption{Memory access times with and without cross-core eviction on the Krait 450 processor. A threshold of 700 clock cycles clearly separates the two timing distributions, which indicates that~\lds~is not present.
  }
  \label{fig:k450_timing}
\end{figure}

To determine the presence of \lds~on the Cortex-A7 and the Krait 450, we execute the cache-timing experiment described in Algorithm~\ref{alg:timing_lockdown}. In addition, timing measurements are used to verify the results obtained for the Cortex-A15, A53, and A57. In all experiments, the \texttt{perf} subsystem of the Linux kernel is used to measure access times with a hardware based clock cycle counter. This was taken from Lipp et al.~\cite{LippEtAl2016}\endnote{The timing measurement code can be retrieved from the GitHub repository at https://github.com/IAIK/armageddon.}.

Figure~\ref{fig:k450_timing} shows the timing data collected with two separate executions of Algorithm~\ref{alg:timing_lockdown} on the Krait 450. The first execution performed the eviction step as defined in the algorithm. The timings measured during the reload phase are shown in red. The second execution skipped the eviction step. These timings are shown in blue, for comparison. When the eviction step is skipped, the target address remains in the cache, and thus the access time is significantly lower, as pictured. This indicates that the Krait 450 does not implement \lds.

\begin{figure}[t!]
  \includegraphics[width=.47\textwidth]{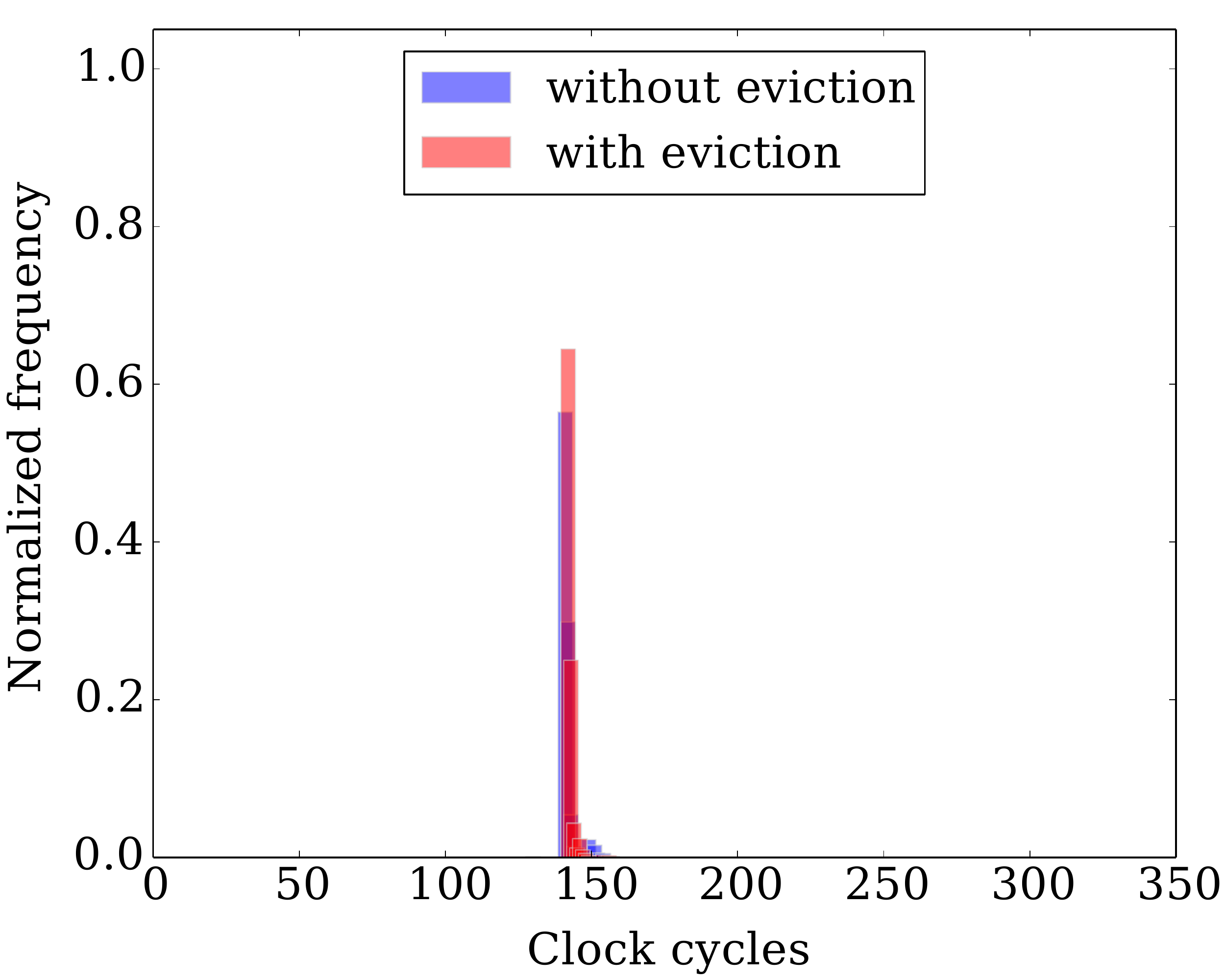}
  \caption{Memory access times with and without cross-core eviction on the ARM Cortex-A57 processor. The two timing distributions clearly overlap, which indicates that~\lds~inhibits the eviction.
  }
  \label{fig:a57_timing}
\end{figure}

The same timing measurements are performed on the Cortex-A57. The results are shown in Figure~\ref{fig:a57_timing}. Since the timings for each execution virtually overlap in the graph, it is clear that the target address is never evicted. This confirms the results for the A57 derived with the hardware debugger, i.e., that it implements \lds. For both the Krait 450 and the A57, we performed 50,000 measurements to ensure that clear trends can be seen.

Corresponding experiments on the Cortex-A7 indicate that its instruction cache side implements \lds. The measurements on the Cortex-A15 and A53 processors confirm the previous results and indicate once more that they implement \lds.

\subsection{Discussion of the Test Results}

\begin{table}[b]
	\centering
	\caption{Evaluation results for the ARM and ARM-compliant processors of the test devices.}
	\label{tab:lockdown_results}
\resizebox{\columnwidth}{!}{%
\begin{tabular}{C{2.25cm}|C{4.75cm}|C{2.25cm}}\hline
	\textbf{Processor} & \textbf{System-on-Chip} & \textbf{\lds} \\
	\hline
	Cortex-A7  & Samsung Exynos 5422      & \textbf{Present}  \\
	Cortex-A15 & Samsung Exynos 5250/5422 & \textbf{Present}  \\
  	Cortex-A53 & ARM Juno r0              & \textbf{Present}  \\
	Cortex-A57 & ARM Juno r0              & \textbf{Present}  \\
	Krait 450  & Qualcomm Snapdragon 805  & Not Present       \\	
	\hline
\end{tabular}
}
\end{table}

The summary of the test results is shown in Table~\ref{tab:lockdown_results}. All ARM Cortex-A processors on our test devices exhibit \lds~in their inclusive last-level caches, whereas no evidence of \lds~can be found on the Qualcomm Krait 450. The practical impact of \lds~is that cache lines in the LLC are transparently locked during runtime. On a multi-core system, this can be triggered simultaneously in multiple cores. For each core, \lds~essentially reduces the number of lines per LLC set that are available to store new data and instructions. Depending on the associativity of the core-private cache levels the LLC is inclusive to, a significant fraction of cache lines can be locked in an LLC set. Referring to Table~\ref{tab:set_assoc}, it can be seen that the Cortex-A7 features 2-way L1 instruction caches and an inclusive 8-way L2 cache. This means that on a quad-core A7 it is possible to lock all ways of an LLC cache set with instructions held in the core-private caches. Requests to store lines from L1 data caches in such a set subsequently fail, but do not violate inclusiveness, as the LLC is non-inclusive to L1 data caches. On the other Cortex-A processors, L2 cache sets cannot fully be locked. In quad-core Cortex-A15, A53, and A57 processors, up to 8 ways can be locked down in an L2 cache set at once. As the Krait 450 does not implement \lds, the number of ways in the LLC does not need to match the sum of ways in the L1 caches. Hence, the L1 data caches contain 4 ways while the LLC contains 8 ways.

\vspace*{1mm}
\section{Related Work}
\label{sec:relwork}
Early practical cache attacks were demonstrated by Bernstein~\cite{Bernstein2005} and Osvik et al.~\cite{OsvikEtAl2006}. While Bernstein based his analysis on the overall execution time of AES encryptions, Osvik et al. implemented two access-driven techniques named \texttt{Evict+Time} and \texttt{Prime+Probe}. These techniques were used to introduce novel attack methodologies and cipher targets. Ac{\i}i\c{c}mez~\cite{Aciicmez2007} and Percival~\cite{Percival2005} utilized \texttt{Prime+Probe} to steal an RSA secret key while Neve and Seifert utilized it to perform an efficient last round attack on AES~\cite{NeveSeifert2007}. Ristenpart et al.~\cite{Ristenpart_hey} used the same technique to recover keystrokes from co-resident virtual machines (VMs) in the Amazon EC2 cloud. This work was later expanded by Zhang et al.~\cite{ZhangEtAl2012}, proving the effectiveness of \texttt{Prime+Probe} to recover El Gamal cryptographic keys across VMs via the L2 cache. Concurrently, a new attack methodology was proposed by Gullasch et al.~\cite{GullaschEtAl2011}, who were able to retrieve an AES key from a core co-resident victim by abusing memory deduplication, the flush instruction \texttt{clflush}, and the Completely Fair Scheduler of Linux. 

Yarom and Falkner~\cite{YaromFalkner2014} extended the work by Gullasch et al. and proposed the \texttt{Flush+Reload} attack, with which they recovered RSA secret keys across processor cores and virtual machines. This work was expanded by Irazoqui et al.~\cite{waitaminute,IrazoquiEtAlAsia2015}, who demonstrated the recovery of AES keys and TLS session messages. The \texttt{Flush+Reload} technique was concurrently used by Benger et al.~\cite{BengerEtAl2014} to recover ECC secret keys, by Zhang et al.~\cite{ZhangEtAl2014} to attack e-commerce applications across PaaS VMs, and by Gruss et al.~\cite{GrussEtAl2015} to implement template attacks. Based on the timing variations of \texttt{clflush}, Gruss et al.~\cite{GrussEtAl2016a} proposed the \texttt{Flush+Flush} attack. As a substitute to the flush instruction, Gruss et al.~\cite{GrussEtAl2015} also proposed the \texttt{Evict+Reload} attack.

\texttt{Flush+Reload}, \texttt{Flush+Flush}, and \texttt{Evict+Reload} require shared memory between an adversary and a target. This requirement was removed by Liu et al.~\cite{LiuEtAl2015} and Irazoqui et al.~\cite{IrazoquiEtAl2015}, who showed the feasibility of \texttt{Prime+Probe} via the last-level cache. Both studies opened a range of scenarios in which cache attacks could be applied. For instance, Oren et al.~\cite{OrenEtAl2015} executed \texttt{Prime+Probe} in JavaScript and Inci et al.~\cite{InciEtAl2016} demonstrated the applicability of the technique in commercial IaaS clouds.

Different aspects of the cache hierarchy have also been exploited to recover sensitive information. Irazoqui et al.~\cite{crossasiaccs} demonstrated the applicability of cache attacks across CPUs through the cache coherency protocol. Yarom et al.~\cite{YaromEtAl2016} showed that cache bank contentions introduce timing variations of accesses to different words on a single cache line. Gruss et al.~\cite{GrussEtAl2016} exploited prefetching instructions to circumvent supervisor mode access prevention and address space layout randomization.

Most of the previous cache attack literature is dedicated to the x86 architecture. Recent works~\cite{LippEtAl2016,ZhangEtAl2016a,ZhangEtAl2016} have made several contributions to overcome the challenges of applying known userspace cache attacks from x86 to ARM processors. \lds~is not recognized or mentioned in any of them. The following section discusses how \lds~relates to these publications and why it might have stayed undetected.

\subsection{{\large \textbf{\lds}} in Previous Work}
\label{sec:comp}
Lipp et al.~\cite{LippEtAl2016} were the first to demonstrate the feasibility of \texttt{Prime+Probe}, \texttt{Flush+Reload}, \texttt{Evict+Reload}, and \texttt{Flush+Flush} attacks on ARM devices. Despite their extensive experiments, the authors did not mention any encounter of a feature similar to \lds. We believe this can mainly be explained with their selection of test devices: the OnePlus One, the Alcatel One Touch Pop 2, and the Samsung Galaxy S6. Respectively, these mobile phones feature the Krait 400, the Cortex-A53, and a big.LITTLE configuration of the Cortex-A53 and Cortex-A57. We assume that the Krait 400, like the Krait 450 we experimented on, does not feature \lds. For the Cortex-A53, we believe the authors relied on evictions caused by background activity on the target core to successfully execute cross-core attacks. They evict instructions from the instruction-inclusive L2 cache using data accesses from a different core, which we have found to be impossible on the Cortex-A53. Indeed, in discussing their cross-core eviction strategy on this processor, they mention ``\textit{the probability that an address is evicted from L1 due system activity is very high}"~\cite{LippMS}, though it is not clear if they are referring to the attacking core, the victim core, or both. The third and final test device, the Samsung Galaxy S6, features a full-hierarchy flush instruction available from userspace by default. Thus, the authors bypassed \lds~ by flushing cache lines instead of evicting them.

Zhang et al.~\cite{ZhangEtAl2016} implemented a variant of the \texttt{Flush+Reload} attack in a zero-permission Android application. The authors also experimented on processors featuring \lds~and did not mention any encounter with it either. In fact, they used the same processors analyzed in this work, namely, the Cortex-A7, A15, A53, A57, and the Krait 450. Since their work was focused solely on \texttt{Flush+Reload}, one of the two cache attacks unaffected by \lds, we assume they simply did not encounter it during their experiments. One of the main contributions of their work was to implement an \emph{instruction-side} \texttt{Reload} in a return-oriented manner, i.e., by executing small blocks of instructions. This contribution stemmed from using the \texttt{cacheflush} syscall in the \texttt{Flush} step, as it only invalidates the instruction side. As a prerequisite for their final target device selection, the authors experimentally determined the inclusiveness property of the last-level caches on all devices. Surprisingly, they concluded that all of the L2 caches in the aforementioned processors are inclusive with respect to the L1 data and instruction caches. This contradicts our experiments, which found the Cortex-A7 and A53 to only be inclusive on the instruction side, and the Cortex-A15 and A57 to only be inclusive on the data side. Further, the official ARM documentation of the Cortex-A7, for example, explains that ``\textit{Data is only allocated to the L2 cache when evicted from the L1 memory system, not when first fetched from the system.}"~\cite{A7TRM}. We understand this to mean that the L2 cache of the Cortex-A7 is \emph{not} inclusive with respect to the L1 data caches. This complies with our observations.

In other previous work, Zhang et al.~\cite{ZhangEtAl2016a} implemented a \texttt{Prime+Probe} attack in an unprivileged Android application on an ARM Cortex-A8. In this work, we did not determine the presence of \lds~on this processor model. However, the test system used by Zhang et al.~\cite{ZhangEtAl2016a} comprised only a single Cortex-A8 processor core. As \lds~does not affect same-core attacks, the experiments of the authors would not have been affected, even if the Cortex-A8 implemented \lds.
\section{Implications of {\large \textbf{\lds}}}
\label{sec:impli}
\ld~prevents the eviction of cache lines from inclusive cache levels, if copies of that line are contained in any of the caches said level is inclusive to. Yet, the ability to evict data and instructions from a target's cache is a key requirement for practical cross-core cache attacks. Table~\ref{tab:blocked_attacks} illustrates the impact of \lds~on state of the art cache attacks implemented on ARM Cortex-A processors. Each row shows one attack technique and the corresponding effect of \lds~in three different scenarios: same-core, cross-core, and cross-CPU. A `\checkmark' or `*'  signifies that an attack can be mounted, whereas a `{\color{red} \xmark}' indicates that \lds~fundamentally interferes with the attack.

\begin{table}[t!]
	\centering
	\caption{Utility of known cache attacks in different scenarios on ARM Cortex-A processors with inclusive caches implementing \lds. `\checkmark' indicates the attack is unaffected by~\lds, while `{\color{red} \xmark}' denotes obstruction by \lds. \texttt{Flush+Reload} and \texttt{Flush+Flush} are uninhibited by \lds~but only apply to a limited number of ARMv8-A SoCs and are thus listed as `*'.}
	\label{tab:blocked_attacks}
	\resizebox{\columnwidth}{!}{
\begin{tabular}{l|c|c|c}\hline
	\multicolumn{1}{c|}{\bf{Attack}} & \bf{Same-core} & \bf{Cross-core} & \bf{Cross-CPU}  \\
	\hline
	Evict\,+\,Time~\cite{OsvikEtAl2006}      & \checkmark              &                             {\color{red} \xmark} &                {\color{red} \xmark}  \\
	Prime\,+\,Probe~\cite{OsvikEtAl2006}     & \checkmark              &                             {\color{red} \xmark} &                {\color{red} \xmark}  \\
	Flush\,+\,Reload~\cite{YaromFalkner2014} & *                       &                                               *  &                                   *  \\
	Evict\,+\,Reload~\cite{GrussEtAl2015}    & \checkmark              &                             {\color{red} \xmark} &                {\color{red} \xmark}  \\
	Flush\,+\,Flush~\cite{GrussEtAl2016a}    & *                       &                                                * &                                   *  \\
	\hline
\end{tabular}}
\end{table}

Given the nature of \ld, all same-core attacks remain possible, as the adversary can evict target memory from all core-private cache levels. All attacks based on a full-hierarchy flush instruction are also not affected by \lds. However, said flush instruction, unlike on x86 processors, is not available on any ARMv7-A compliant processor and must be enabled in control registers on ARMv8-A compliant processors. Access to these control registers is limited to privileged, i.e., kernel or hypervisor code. These flush based attacks, namely \texttt{Flush+Reload} and \texttt{Flush+Flush}, are therefore denoted with `*'.

All attacks based on evicting a cache line by accessing set-congruent addresses are affected by \ld~in cross-core scenarios. \texttt{Evict+Time}, \texttt{Prime+Probe}, and \texttt{Evict+Reload} techniques are all impaired by \lds. To leverage their full potential, circumvention strategies must be found and employed. In general, such strategies can also be used to target non-inclusive LLCs, where cross-core evictions are not possible, either. In the upcoming section, we discuss circumvention strategies and demonstrate that the attack proposed by Irazoqui et al.~\cite{waitaminute} can still be mounted in a cross-core \texttt{Evict+Reload} scenario with an inclusive LLC implementing \lds.

\section{Circumventing {\large \textbf{\lds}}}\label{sec:circum}
Despite the restrictions that \ld\ poses to eviction based cache attacks, its effects can be alleviated with the following strategies:

\begin{itemize}
  \item {\bf Pre-select Target SoCs:} Our findings suggest that \lds~is present on Cortex-A cores designed by ARM itself, while it is not implemented by ARM compliant cores, such as Qualcomm's Krait 450. As the concept is protected by US patent~\cite{Williamson2012}, ARMv7-A and ARMv8-A compliant cores would have to pay royalties to implement \lds. 
By exclusively targeting Cortex-A compliant processors not implemented by ARM, chances of not encountering \lds~increase. Alternatively, \texttt{Flush+Reload} or \texttt{Flush+Flush} based attacks can still be mounted on ARMv8-A SoCs that offer the cache flush instruction in userspace, i.e. the Samsung Galaxy S6~\cite{LippEtAl2016}.

  \item{\bf Achieve Same-core Scenario:} Certain attack scenarios realistically allow the adversary to execute code on the same core as the target program. Since same-core attacks are not affected by \lds, this entirely removes its impact. ARM TrustZone, e.g., enables secure execution of trusted code on an untrusted operating system. Given that the untrusted OS is compromised by the adversary, the trusted code can be scheduled to run on any given processor core. By matching the core affinity of the attacking process to the one of the respective trusted target, the attack is reduced to a same-core setting and can successfully be mounted, even across TEE boundaries~\cite{ZhangEtAl2016a}. 

  \item{\bf Trigger Self-evictions:} When \lds~is active, a cache line can only be evicted from the inclusive LLC if no higher cache level contains a copy of it. If  the target program offers services to the rest of the system, the adversary can try to issue requests such that the core-private cache of the target is sufficiently polluted and the cache line under attack is evicted. The target program essentially performs a `self-eviction' and thus re-enables LLC evictions and consequently cross-core attacks.

  \item{\bf Increase Load and Waiting Time:} Inclusive caches with \lds~require that the number of ways in lower levels are greater or equal than the sum of ways in all higher cache levels. This limits the associativity of core-private caches, which the LLC is inclusive to, especially on multi-core systems. If the attack allows, an adversary can take advantage of the low associativity and simply prolong the waiting time between reloads such that the target line will automatically be evicted from core-private caches by other system activity scheduled on the respective core. To amplify the effect, the adversary can also try to increase overall system load, e.g., by issuing requests to the OS aiming at increasing background activity in the targeted core.

  \item{\bf Target Large Data Structures:} Self-evictions, high system load, and prolonged waiting times all increase the chances that a cache line is evicted by itself from core-private caches. The success rate of an attack is further improved, if multiple cache lines can be targeted. The more lines that are exploitable, the higher the chances that at least one of them is automatically evicted from core-private caches. For example, the transformation tables (T-tables) of AES software implementations span multiple cache lines, i.e., 16 lines for a 1\,kiB table with 64 bytes per line. The attack proposed by Irazoqui et al.~\cite{waitaminute} observes the first line per table to recover an entire AES key. The authors note that ``\textit{any memory line of the T table would work equally well.}'' In the upcoming section, we pick up this idea and demonstrate how the attack can be extended to exploit multiple cache lines to successfully circumvent \lds.
\end{itemize}

Note that all of the presented strategies increase the chances of successful attacks not only on inclusive caches implementing \lds, but also on \emph{non-inclusive} caches.

\subsection{Attack on AES}\label{sec:circum:subsec:aes}
Irazoqui et al.~\cite{waitaminute} propose an attack on table based implementations of AES using \texttt{Flush+Reload}. The basic strategy is to flush one cache line per table before an encryption and reload it afterwards. If any lookup table value stored on the observed cache line is used during encryption, the adversary will encounter fast reload times. If said line is not accessed, it will be fetched from memory and reload times will be slow. With all table lookups dependent on the secret key, the adversary can infer bits of the key from the observed reload times.

In table based implementations of AES, each cache line has a certain probability with which it is \textit{not} used during en- or decryption. This probability depends on the size and the number of the tables as well as the size of the cache lines. It is defined as

\begin{equation}
  P_{na} = \left(1-\frac{t}{256}\right)^{n}\,\,.
  \label{eq:probaes}
\end{equation}

Variable $t$ denotes the number of table entries stored on a cache line. For 4-byte entries and a 64-byte cache line, $t = 16$. Variable $n$ defines the number of accesses to the table that a cache line is part of. Given an AES-128 implementation that uses four 1\,kiB T-tables and performs 160 lookups per encryption, which evenly spread over the four tables, $n = 40$. With $t = 16$, this yields a no-access probability of $P_{na} = 0.0757$. Note that the attack exploits observations of not accessed cache lines. As a result, the number of required observations increases, as $P_{na}$ gets smaller.

The attack targets the last round of AES, i.e., the 10$^{th}$ round. It is shown in Equation~\ref{eq:lastround}. The ciphertext is denoted as $c_{i}$ ($i = 0..15$). The 10$^{th}$ round key is given as $k_{i}^{10}$, whereas the state of AES is defined as $s_{i}$. The target lookup table used in the last round is denoted as $T$. For each encryption, the adversary keeps track of the reload times and the ciphertext. If successful, the attack recovers the last round key. For the recovery phase, the last round is re-written as

\begin{equation}
  c_{i} = k_{i}^{10} \oplus T\left[s_{i}\right] \,\,\,\rightarrow\,\,\, k_{i}^{10} = c_{i} \oplus T\left[s_{i}\right]\,\,.
  \label{eq:lastround}
\end{equation}

\paragraph{Improvements} The original attack targets one cache line per table. If the observations of said line are of poor quality, the attack is prolonged or fails. This can happen on a processor that implements \lds\ or non-inclusive caches. If LLC evictions fail, the adversary cannot determine whether the selected cache line has been used during encryption. Irazoqui et al.~\cite{waitaminute} state that the attack works equally well with all cache lines carrying lookup table entries. As discussed in the previous section, it is likely in practice that some of them are automatically evicted from core-private caches, hence re-enabling the attack despite \lds. To leverage observations from all available cache lines, we propose three improvements to the original attack:

\begin{enumerate}
  \item \textbf{Majority vote:} All available cache lines $l$ are attacked ($l = 0..L$). This yields $L$ recovered keys. For each key byte, a majority vote is done over all $L$ recovered values. The value with the most frequent occurrence is assumed to be the correct one. If two or more values occur equally frequently, the lowest one is chosen. The majority vote ensures that wrong hypotheses from noisy cache lines are compensated for as long as the majority of lines yield correct results.

  \item \textbf{Probability filter:} The reload times allow to calculate the actual no-access probability for each cache line, $\tilde{P}_{na}^{l}$. For each table, the line closest to the expected theoretic probability, $P_{na}$, is taken and used in the attack. Lines showing distorted usage statistics due to noise and interference are discarded.

  \item \textbf{Weighted counting:} Every cache line is assigned an individual score $S_{l}$ that is counted each time a key byte hypothesis is derived from the line's reload times. The score is based on the absolute difference of the no-access probabilities, $d_{l} = abs\left(P_{na} - \tilde{P}_{na}^{l}\right)$. It is defined as $S_{l} = 1 - \frac{d_{l}}{1 - P_{na}}$. After all scores have been added for all hypotheses, the recovery phase proceeds as proposed.
\end{enumerate}

\begin{figure}[t!]
	\centering                            
    \includegraphics[width=0.47\textwidth, trim=0cm 0cm 0cm 0cm, clip]{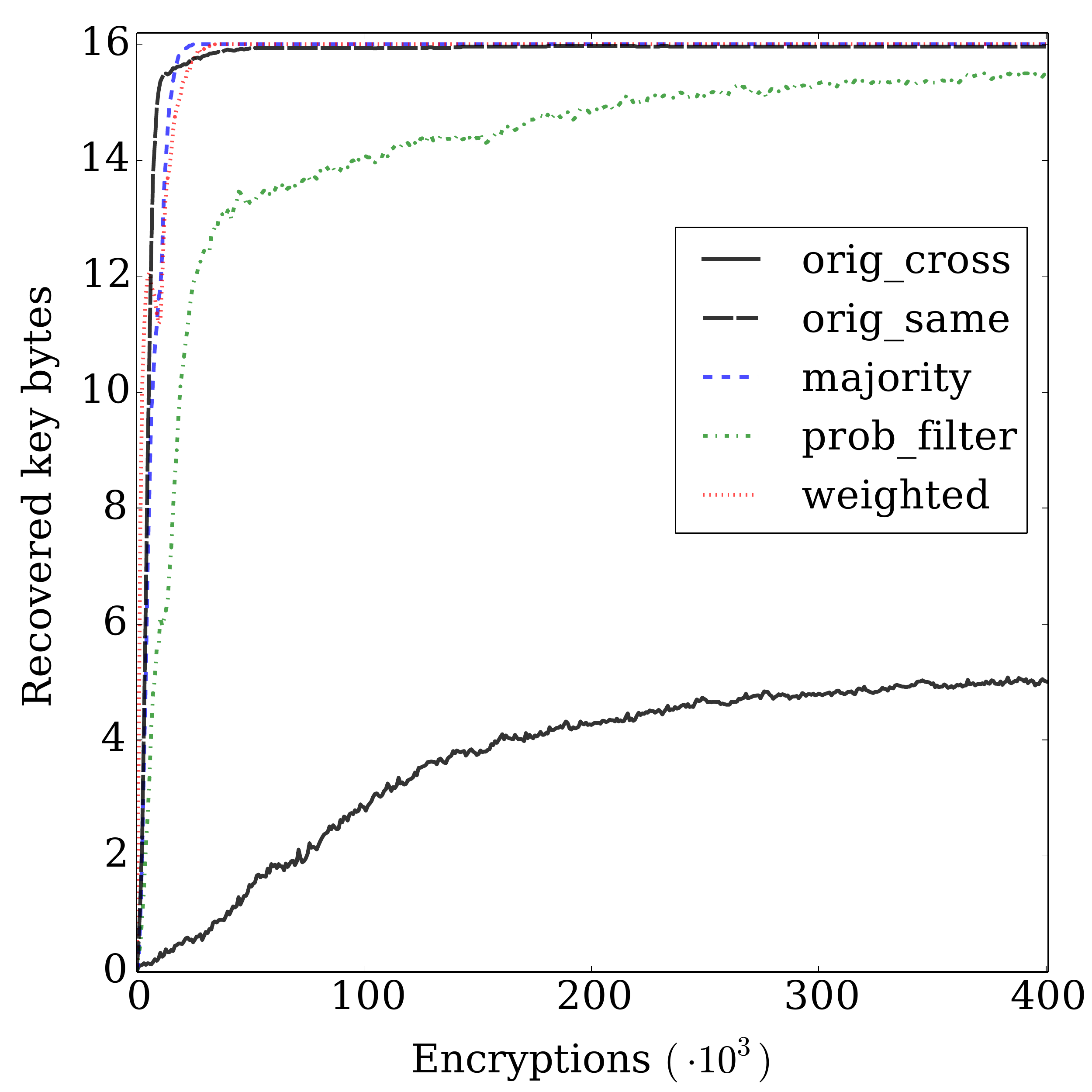}
    \caption{\texttt{Evict+Reload} attacks on an ARM Cortex-A15 with \lds~performed with 100 random keys. The number of key bytes recovered on average are displayed for an increasing number of encryptions.}
    \label{fig:a15attacks}
    \vspace*{5mm}
\end{figure}

We implement the original attack and all improvements using \texttt{Evict+Reload} on a multi-core ARM Cortex-A15 processor featuring a data-inclusive LLC with \lds. We employ sliding window eviction with parameters 36-6-2 (see Table~\ref{tab:set_assoc}). The targeted AES implementation uses four 1\,kiB T-tables. The adversary and target processes are running on top of a full-scale Linux operating system. In total, we perform five different attacks. First, we implement the original attack with adversary and target on the same core and then on separate cores. This illustrates the impact of \lds, which only affects cross-core attacks. The rest of the attacks are conducted with adversary and target on separate cores and each demonstrate one of the proposed improvements. The results are illustrated in Figure~\ref{fig:a15attacks}. Each attack is repeated for 100 random keys and the average number of correctly recovered key bytes is shown over an increasing number of encryptions. It can clearly be seen that \lds~impairs the original cross-core attack (\textit{orig\_cross}). After 400,000 encryptions no more than 5 key bytes are correctly recovered. The fact that at least some key bytes are correct is owed to sporadic and automatic evictions of the observed cache lines from the target's core-private cache. These evictions can be caused by stack and heap data accesses (such as AES state and key arrays as well as their pointers), and possibly by unrelated processes scheduled on the same core. Attacks in the same-core setting (\textit{orig\_same}) are not affected by \lds~and allow full-key recovery. Our improvements clearly demonstrate that cross-core attacks are still possible, if multiple cache lines can be observed. Both majority vote (\textit{majority}) and weighted counting (\textit{weighted}) recover the entire key with less than 100,000 encryptions and therefore offer similar success rates as the same-core attack. The probability filter (\textit{prob\_filter}) still allows full-key recovery within 100,000 encryptions, if a brute-force search with complexity $<\,2^{32}$ is added.

The results illustrate that even on processors implementing \lds~cache attacks can still be successful in practice, if multiple cache lines are monitored. Note that the proposed improvements are also beneficial on processors without \lds\ or on systems with non-inclusive caches. If attacks rely on observing a specific cache line, the chances of success are significantly reduced on processors implementing \lds.

\section{Conclusion}
\label{sec:concl}
The licensing ecosystem of ARM drives an increasingly heterogeneous market of processors with significant microarchitectural differences. Paired with a limited understanding of how ARM's cache architectures function internally, this makes assessing the practical threat of flush and eviction based cache attacks on ARM a challenging task. Although the feasibility of state of the art attacks has been demonstrated, their requirements are far from being fulfilled on all ARM processors. Flush instructions are supported only by the latest architecture version and must explicitly be enabled for userspace applications. This limits the practical utility of flush based attacks. Last-level caches can be non-inclusive, impeding practical cross-core eviction attacks that require LLCs to be inclusive. Our work shows that these attacks can still fail even on inclusive LLCs, if \lds~is implemented. On the contrary, more sophisticated attack techniques seem to overcome both \lds~and non-inclusive cache hierarchies. We therefore believe that a fair and comprehensive assessment of ARM's security against cache attacks requires a better understanding of the implemented cache architectures as well as rigorous testing across a broad range of ARM and thereto compliant processors.

%
%
{\footnotesize \bibliographystyle{acm}
\bibliography{references}}

%
%
\theendnotes

\end{document}